\documentclass[aOAps,superscriptaddress,fullpaper,twocolumn,footinbib]{revtex4-1}
\usepackage[dvipdfmx]{graphicx}
\usepackage{amsmath}
\usepackage{amssymb}
\usepackage{braket}
\usepackage{bm}
\usepackage{color}
\usepackage{verbatim}

\begin{document}

\title{
Phase diagram of disordered higher-order topological insulator: \\A machine learning study
}
\author{Hiromu Araki}
\email{araki@rhodia.ph.tsukuba.ac.jp}
\affiliation{Graduate School of Pure and Applied Sciences, University of Tsukuba, Tsukuba, Ibaraki 305-8571, Japan}
\author{Tomonari Mizoguchi}
\affiliation{Department of Physics, University of Tsukuba, Tsukuba, Ibaraki 305-8571, Japan}
\author{Yasuhiro Hatsugai}
\affiliation{Graduate School of Pure and Applied Sciences, University of Tsukuba, Tsukuba, Ibaraki 305-8571, Japan}
\affiliation{Department of Physics, University of Tsukuba, Tsukuba, Ibaraki 305-8571, Japan}
\begin{abstract}
A higher-order topological insulator is a new concept of topological states of matter,
which is characterized by the emergent boundary states whose dimensionality is lower by more than two 
compared with that of the bulk,
and draws a considerable interest. 
Yet, its robustness against disorders is still unclear.
In this work, we investigate a phase diagram of higher-order topological insulator 
phases in a breathing kagome model 
in the presence of disorders, 
by using a state-of-the-art machine learning technique.  
We find that the corner states survive against the finite strength of disorder potential as long as the energy gap
is not closed, indicating the stability of the higher-order topological phases against the disorders.

\end{abstract}

\maketitle
\section{Introduction}
Understanding and classifying topological states of matter are central issues in today's condensed matter physics~\cite{RevModPhys.82.3045,RevModPhys.83.1057,RevModPhys.89.041004}.
Among a number of such systems, topological states in gapped free-fermion systems are the most well-understood example. 
There, the keen interplay between symmetry and topology gives rise to various non trivial states, 
and the elegant mathematical tools such as group theory, Clifford algebra, and K-theory allow us to classify topological insulators (TIs) and superconductors (TSCs)
under the internal~\cite{
PhysRevB.78.195125,doi:10.1063/1.3149495,1367-2630-12-6-065010}, space-group~\cite{Slager2012,PhysRevB.88.125129,Po2017} 
and magnetic space-group~\cite{Watanabeeaat8685} symmetries. 

However, this is not the end of the story. 
Very recently, a new class of topological insulators has been found, which is called 
a higher-order topological insulator (HOTI)~\cite{Hayashi2018, PhysRevB.95.165443, PhysRevB.96.245115, Benalcazar61, PhysRevLett.119.246402, Schindlereaat0346,PhysRevLett.120.026801,PhysRevB.97.241405, Xu2017}.
In this newly-introduced concept, the ``order'' means the dimensionality of the boundary state. 
Namely, in conventional TIs in $d$-dimensions, the gapless boundary states emerge at the $(d-1)$-dimensional boundary
~\cite{PhysRevLett.71.3697,PhysRevB.48.11851}, and in that sense, conventional TIs are the first-order TIs. 
Similarly, in $n$th order TIs ($2 \leq n \leq d $), 
gapless boundary states emerge at the $(d-n)$-dimensional boundary.
In many cases, the emergence of such gapless states is protected by the spacial symmetries 
such as mirror symmetries and rotational symmetries. 
\begin{figure}[b]
 \begin{center}
 \includegraphics[width=\linewidth]{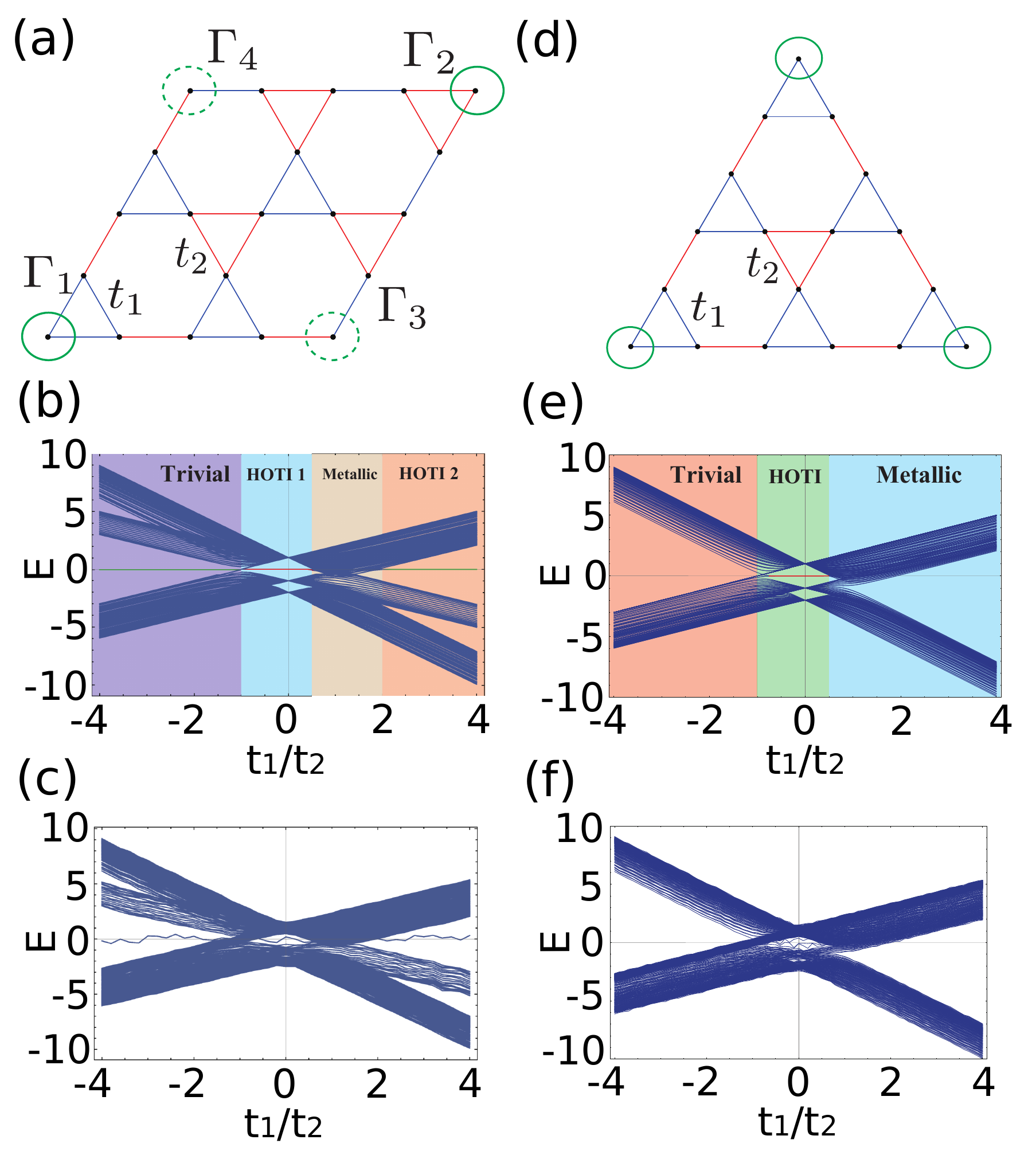}
 \caption{
 A kagome flake with (a) a rhombus geometry and (d) a triangle geometry. 
In each geometry, corners are denoted by green circles. 
The energy spectrum of the rhombus as a function of $t_1/t_2$ 
for (b) a rhombus geometry and (e) a triangle geometry. 
The energy spectrum with the disorder $W=1.1$ for (c) a rhombus geometry and (f) a triangle geometry. 
  }
  \label{fig:1}
 \end{center}
\end{figure}

To reveal the physical properties of HOTIs, studying the effects of disorders is important. 
Indeed, for the first-order TIs protected by the time-reversal symmetry, 
the most essential property is the prohibition of back-scatterings of the helical edge/surface states. 
As a results, the phase diagrams of the disordered TIs can be obtained by investigating the conductance~\cite{PhysRevB.79.045321,1367-2630-12-6-065008,PhysRevB.85.155138,PhysRevLett.110.236803},
and their validity is evidenced by the direct calculation of topological indices~\cite{PhysRevB.31.3372,PhysRevB.83.195119,doi:10.1063/1.4942494,doi:10.1063/1.5026964,doi:10.7566/JPSJ.86.123710}. 
For HOTIs, on the other hand,
the definition of the disordered HOTIs,
relying on either physical quantities or topological indices, 
is still unclear. 
Nevertheless, we expect that 
the boundary states of the HOTI are localized even in the presence of weak disorders, 
and that the existence of the boundary states may serve as an indicator of the HOTI phase in the disordered systems.
To examine this scenario, it is desirable to employ a method that can distinguish the disordered HOTI from other phases 
in a systematic manner.

To this end, in this paper, we investigate the robustness of the HOTI against disorders by using a machine learning method. 
The machine learning method, especially the neural network,
is widely used in the various fields in physics,
including quantum many-body systems~\cite{Carleo602,PhysRevB.96.205152,PhysRevB.97.075114, PhysRevB.97.134109, 
Huembeli2018}, 
Monte Carlo simulations~\cite{PhysRevB.97.205140},
high-energy physics~\cite{WHITESON20091203,Baldi2014} and astrophysics~\cite{doi:10.1093/pasj/psw096}. 
As for the topological states, the first order TIs and TSCs
have been successfully classified in the presence of disorders,
and the resulting phase diagrams reproduce those obtained by the other methods.
~\cite{doi:10.7566/JPSJ.85.123706, doi:10.7566/JPSJ.86.044708,PhysRevB.97.205110}.
This motivates us to apply this method to disordered HOTIs. 

In the present work, we employ a tight-binding Hamiltonian on a breathing kagome lattice~\cite{0295-5075-95-2-20003,PhysRevB.97.241405,PhysRevLett.120.026801, Xu2017}. 
The Hamiltonian, as we will explain later, is known to possess the second-order TI phases 
with zero-dimensional corner state
protected by symmetries~\cite{PhysRevB.97.241405,PhysRevLett.120.026801, Xu2017}.
For this system, we first perform supervised learning in the clean limit, where
the phase diagram has already been known~\cite{PhysRevB.97.241405,PhysRevLett.120.026801, Xu2017}. 
Once the supervised ``model'' for classification is obtained, 
we then introduce on-site disorders and classify the phases by using that model. 
We obtain phase diagrams for the disordered breathing kagome model
in both rhombus and triangle geometries.
For both of them, we find that the HOTI phase survives as far as the energy gap does not collapse,
due to the robustness of the corner states against the disorders.

The rest of this paper is organized as follows.
In Sec. \ref{sec:model}, we introduce our model and review the phase diagrams in the clean limit. 
The symmetries of Hamiltonian and the topological invariants proposed in the previous works are also summarized.
In Sec. \ref{sec:method}, we explain our method to identify the phases by using the supervised learning. 
In Sec. \ref{sec:result}, we present our main results, namely, the phase diagrams in the presence of disorders. 
In Sec. \ref{sec:summary}, we present a summary and discussions.
The comparison with the obtained results of the inverse participation ratio is discussed in Appendix.

\section{Model \label{sec:model}}
We consider a tight binding Hamiltonian for spinless fermions on a breathing kagome lattice:
\begin{equation}
 \label{eq:H_0}
  \mathcal{H}_0 = -\sum_{i,j} t_{ij} c_i^\dagger c_j
\end{equation}
where $c_i^\dagger$ and $c_i$ are, respectively, the creation and annihilation operators of an electron on
a site $i$, and
$t_{ij} = t_1 (t_2)$ if the bond between $i$ and $j$
belongs to the nearest-neighbor bond and 
it lives on the upward 
(downward) triangles [Fig. \ref{fig:1}(a) and \ref{fig:1}(d)]. 
In the following, we take $t_2$ to be positive.

The band structure of the bulk is obtained by diagonalizing $ \mathcal{H}_0$
in the Fourier space. 
There exists a flat band with energy $t_1 + t_2$.
The other two bands are gapped when $|t_1| \neq |t_2|$;
the gapless linear dispersion appears at $K$ and $K'$ for $t_2 =t_1$,
and at $\Gamma$ point for $t_2 = -t_1 $

\subsection{Phase diagrams in the clean limit for rhombus and triangular geometries}
To study HOTI phases, we have to consider finite system with an open boundary condition. 
So far, it is known that there are two choices of global geometries, namely, rhombus and triangular geometries
[see Fig. \ref{fig:1}(a) and \ref{fig:1}(d), respectively]. 
In both of these geometries, the model has HOTI phases with zero-dimensional 
localized states, i.e. corner states~\cite{PhysRevLett.120.026801, Xu2017,PhysRevB.97.241405}. 
This can be understood by considering the limit of $|t_1| = 0$ or $|t_2| = 0 $. 
For instance, if  $|t_1|  = 0 $, the \lq \lq trimers" are formed on all downward triangles,
and the \lq \lq dimers" at all edges which do not belong to the downward triangles. 
For a rhombus geometry, only the site at $\Gamma_1$ corner is isolated and serves as a zero-energy mode,
and it survives even for finite $t_1$, as long as $|t_1| < |t_2|$ is satisfied. 
Similarly, for $|t_1|  > |t_2|$, the corner state appears at $\Gamma_2$ corner.
Indeed, in Ref. \onlinecite{PhysRevB.97.241405}, 
it was shown that the corner zero modes can be explicitly constructed when $|t_1|  \neq  |t_2|$. 
Note that the other two corners, $\Gamma_3$ and $\Gamma_4$, do not possess corner states at any parameters. 
For a triangular geometry, on the other hand, all three corners belong to upward triangles, so the corner states exist
only when $ -1 \leq t_1 /t_2 \leq 1/2 $, 
and those corner states have three-fold degeneracy~\cite{PhysRevLett.120.026801}.

The difference of corner states with different geometry is also crucial when defining the phases.
In a rhombus geometry, the corner state exists for every parameter, but it matters whether the corner state are in-gap state or not. 
So, we classify the phases in such a way that we fix the electron number as $N = \nu_{\rm R} = M^2$, where we consider the system contains 
$M \times M$ rhombuses ($3M^2 - 2M$ sites), and see what the highest occupied state is. 
With $t_1/t_2$ as a tuning parameter, there are four phases~\cite{Xu2017} [Fig. \ref{fig:1}(b)].
For $t_1/t_2 > 2$, the system is in a HOTI phase, in which a zero-energy state localized at 
the $\Gamma_2$ corner exists in a band gap.
We referred to this state as ``HOTI 2" named after the position of the corner state. 
For $1/2 < t_1/t_2 < 2$, the system is in a metallic phase, which has zero-energy bulk or edge states,
and the corner state is buried into those states.
For $-1 < t_1/t_2 < 1/2$, there is another HOTI phase, in which a zero-energy state localized at 
the $\Gamma_1$ corner and it is in the band gap.
We label this state ``HOTI 1".
For $t_1/t_2 < -1$, the system is in a ``trivial'' phase. 
In this phase, there is
a zero-energy state localized at the $\Gamma_2$ corner in the gap, but 
it is not the highest occupied state. 
Rather, one of the degenerated flat modes is the highest occupied state and it masks the corner state. 

In contrast to a rhombus geometry, there are only three phases for a triangle geometry 
since there is only one HOTI phase for $ -1 \leq t_1 /t_2 \leq 1/2 $~\cite{PhysRevLett.120.026801}.
In this phase,
the corner states appear at $(\nu_{\rm T}-1)$-th, $\nu_{\rm T}$-th, and $(\nu_{\rm T}+1)$-th states, with $\nu_{\rm T} = \frac{M(M+3)}{2}$,
for the system consists of $\frac{M(M-1)}{2}$ upward triangles [i.e. $\frac{3 M(M-1)}{2}$ sites]. 
We therefore fix the electron number as $N =\nu_{\rm T}$ so that the highest occupied state is the corner state in this region.
The other two phases are trivial ($ t_1 /t_2 \leq-1$) and metallic ($ t_1 /t_2 \geq 1/2 $) phases, 
which are essentially the same as those for the rhombus geometry.  

\subsection{Symmetries and topological invariants}
So far, it was revealed that zero-energy corner state in the HOTI phase is protected by the symmetries of the Hamiltonian. 
However, the protecting symmetries and corresponding topological invariants vary according to the geometries of the system. 
For the triangle geometry, the gapless corner state is associated with 
the distance of the ``Wannier center'' from the origin, $P_3$, which
is quantized as $1/2$ for the HOTI phase and $0$ for the non-topological phases
\cite{PhysRevLett.120.026801}.
The quantization of $P_3$ originates from the combination of the mirror and three-fold rotational ($C_3$) symmetries. 
For the rhombus geometry, the topological invariant can be obtained from the trajectory of the eigenvalues of the ``Wannier Hamiltonian''
during the adiabatic deformation of the Hamiltonian~\cite{Xu2017}. 
In this case, the emergence of the gapless corner state is associated with the composite symmetry operation of the three-fold rotation and the complex conjugation.

Importantly, for both of these two, 
the spacial symmetries of the Hamiltonian plays a crucial role, while the random impurity potential breaks these symmetries. 
Hence, the topological invariants discussed above are not quantized in the disordered systems.
Also, the corner state, if exists, is no longer pinned at the zero-energy [see Figs.~\ref{fig:1}(c) and \ref{fig:1}(f)].
In such a situation, the definition of the HOTI phase itself is more or less subtle, but 
it may be reasonable to adopt the existence of the corner states as a working definition of the (disordered) HOTI phase. 
This is the reason why we employ a machine learning method, by which we can systematically judge whether the corner states exist as an in-gap state.

The alternative approach to define the HOTI phase is to use the topological invariant which is well-defined even in the absence of the translational symmetry. 
In this respect, the $\mathbb{Z}_3$ Berry phase~\cite{0295-5075-95-2-20003}, which is defined for the local gauge twist of the Hamiltonian, will be a candidate. 
The relation between the $\mathbb{Z}_3$ Berry phase and the HOTI phases will be discussed elsewhere~\footnote{H. Araki, T. Mizoguchi, and Y. Hatsugai, in preparation.}.

\subsection{Effect of disorders}
To study the effects of disorders, we introduce a on-site random potential, $\mathcal{H}_R$,
where 
\begin{eqnarray}
\label{eq:H_R}
\mathcal{H}_R =  \sum_{i} w_i c_i^\dagger c_i.
\end{eqnarray}
Here $w_i$ are randomness with a uniform distribution in $[-W/2, W/2)$ with fixed $W$.
The total Hamiltonian we consider is $\mathcal{H} = \mathcal{H}_0 + \mathcal{H}_R$. 

\section{Identification of phases by machine learning \label{sec:method}}
We classify the phases of the disordered system 
by using a machine learning technique. 
Here a neural-network-based algorithm is implemented by 
using the open-source library PyTorch~\cite{pytorch}. 

\begin{figure*}[htb]
 \begin{minipage}[t]{0.9\textwidth}
  \centering 
  \includegraphics[width=1.0\linewidth]{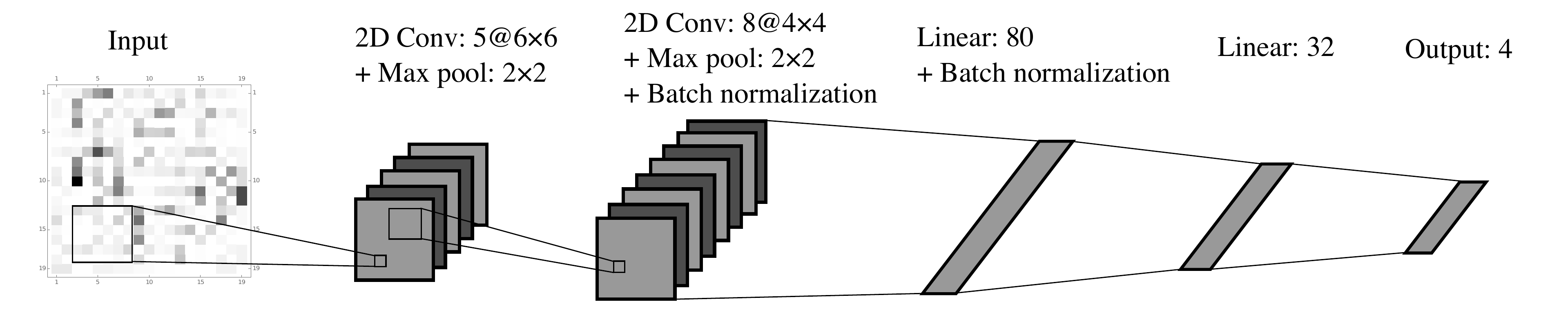}
  \caption{The architecture of the neural network. The first two layers are two-dimensional 
  convolutional layers and the rests are linear layers. 
  We use some max pooling layers and 
  batch normalization. The size of the layers are shown in the figure.}
  \label{fig:arc}
 \end{minipage}
\end{figure*}

\subsection{Architecture of the neural network}
A schematic picture of the architecture of the neural network used in this study is shown in Fig. \ref{fig:arc}. 
We use the convolutional neural network with five layers, in which the first two layers are
two dimensional convolutional layers and the rests are linear layers.
All the activation functions are the ReLU functions $f(x)=\max(0, x)$.
After the first convolutional layer, we use the max pooling layer. 
After the second convolutional layer, 
we use the max pooling layer and the batch normalization.
The max pooling layer takes the max value in $2\times 2$ sectors.
Next layer is a linear layer. We use the batch normalization for it.
Then the neural network outputs the probabilities of phases. 
The number of output is four for rhombus geometry and three for triangle geometry, 
corresponding the number of phases.

\subsection{Input and output data}
Our input data is the electron density of the highest occupied single-particle state.
This choice is similar to the previous works on the first order TIs~\cite{doi:10.7566/JPSJ.85.123706,PhysRevB.97.205110}. 
For a rhombus geometry, the input data is
$|\phi_{i}^{\nu_{\rm R} }|^2$ defined as $c_{\nu_{\rm R}} = \sum_{i} \phi_{i}^{\nu_{\rm R}}c_{i}$,
with $c_{\nu_{\rm R}}$ being the annihilation operator of the state $\nu_{\rm R}$. 
For a triangle geometry, the input data is $( |\phi_{i}^{\nu_{\rm T}-1 }|^2 +|\phi_{i}^{\nu_{\rm T} }|^2 + |\phi_{i}^{\nu_{\rm T}+1 }|^2) /3$, 
which is the average electron density among three corner states. 
This particular choice is convenient since it preserves the three-fold rotational symmetry inherent in this geometry. 
We remark that we employ the present choice of input data because it is suitable for detecting the corner state in the HOTI phase. 
There are other possible choices of the input data, such as the all of the occupied states~\cite{PhysRevB.97.134109}
and topological numbers~\cite{PhysRevB.98.085402}.

\begin{figure}[htb]
 \begin{minipage}[t]{0.5\textwidth}
  \centering
  \includegraphics[width=1.0\linewidth]{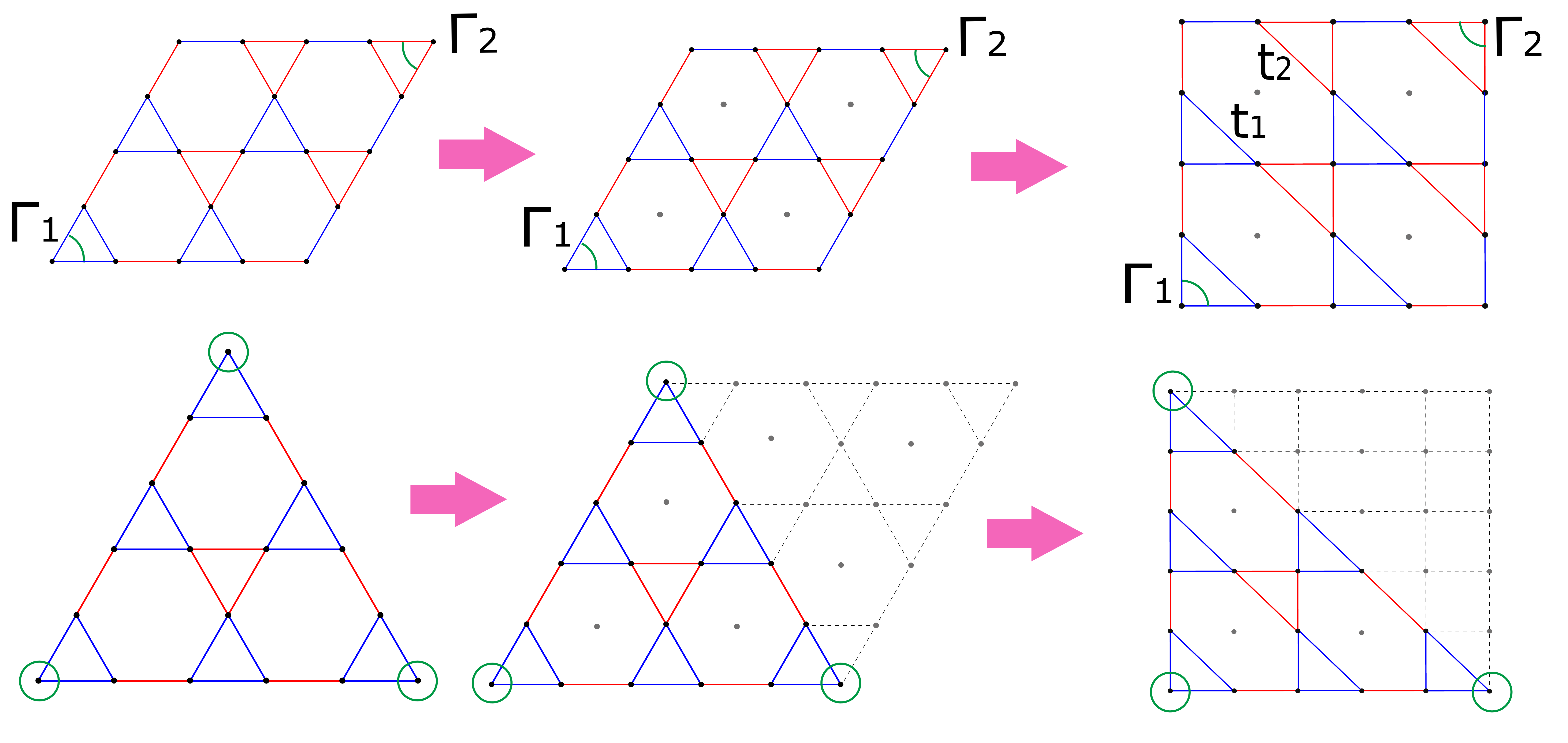}
  \caption{The modification of the kagome lattice to the square lattice
  for a rhombus geometry (upper) and a triangle geometry (lower).
  We add empty sites denoted by gray circles. }
  \label{fig:alignment}
 \end{minipage}
\end{figure}

To make the input data as two-dimensional square-shaped images, we transform the geometry from a kagome lattice to a square lattice with vacancy sites
 (Fig. \ref{fig:alignment}).
For a rhombus geometry, this is achieved by adding redundant sites (denoted by gray circles)
at the center of hexagonal plaquettes, and deform 
equilateral triangles into right triangles. 
Then we make input data from a single-particle wave function, in which probabilities on the added sites
are always zero.
Through this procedure, the size of image data is increased from about $3N$ to $4N$, where $N$ is the 
number of unit cells.
Although the computational cost is increased, we do not lose any information during this process.
Similarly, for a triangle geometry, we add the redundant sites at the center of the plaquettes. 
In addition, we also need to add a large downward triangle composed of the redundant sites.
Then, in the resulting image, more than half of the sites are always empty.  
The output data are in the form of four- (three-) dimensional vectors for a rhombus (triangular) geometry whose 
$n$th component is interpreted as a probability that the state belongs to $n$th phase.

\subsection{Supervised learning}
To classify the phases, we first perform a supervised learning to create a model. 
To do this, we first prepare $5000$ images with randomly-chosen values of $t_1/t_2$ in the clean limit.
Each image is tied with the label of the phases which is represented by 
a four- (three-) dimensional vector for a rhombus (triangular) geometry.
For a vector corresponding to the $n$-th phase,
only $n$-th component is unity and all the others are zero, 
namely, we employ a one-hot representation of the phases.  
Then the model is trained such that the deviation from the answer and the output (or a cost function) is minimized. 
This can be achieved by updating the parameters in the model on the basis of a gradient method. 
Note that 
In the present system, the obtained model can reproduce the data over 99\% of accuracy.

Then, using the trained model, we identify the phases of the disordered system.
To do this, we input the wave functions of the disordered system to the model, and see the output vectors which predict the probability for each phase.
For a given parameters, $t_1/t_2$ and $W$, 
we prepare $30$ samples and investigate the phase diagram in two different methods of averaging these samples. 
The first method is to take an average for the input data. 
Namely, the input is chosen as the averaged wave functions. 
As is pointed out in Ref. \onlinecite{PhysRevB.97.205110}, this procedure is necessary to restore the translational symmetry that is broken by disorders.
The second method is to take an average for the output data.
Namely, we input the single-shot wave functions for $30$ samples and take an average over the obtained output vectors.  
These two methods complement each other. 
To be more specific, the former method is advantageous to distinguish the corner states 
from the randomly-localized states for large $W$, because the randomly-localized states have uniform distributions after taking the average.
However, this method can not distinguish the trivial, metallic and randomly-localized states, 
because all of these states have uniform distributions.
On the other hand, the latter method can distinguish them, because the difference between them is clear in the single-shot wave functions.

\begin{figure*}[t]

 \begin{center}
  \includegraphics[width=0.98\linewidth]{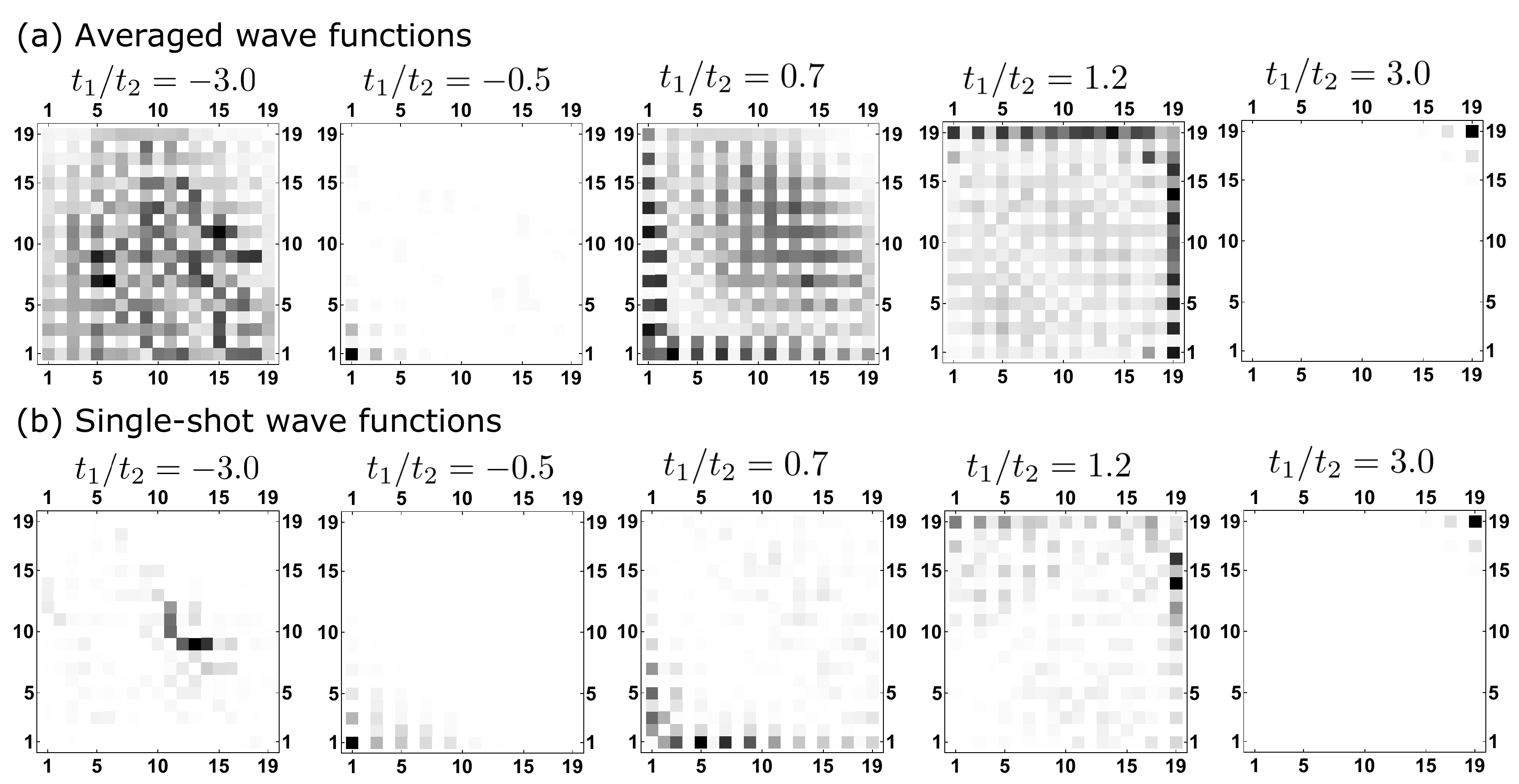}
\caption{
  Wave functions on a rhombus geometry.
  (a) The averaged wave functions of the highest occupied state 
  with $W=1$.
  The average is taken over 30 samples.
  (b)
  The single-shot wave functions of the highest occupied state 
  with $W=1$.}
  \label{fig:states}
 \end{center}
\end{figure*}

\begin{figure*}[t]
 \begin{center}
  \includegraphics[width=0.98\linewidth]{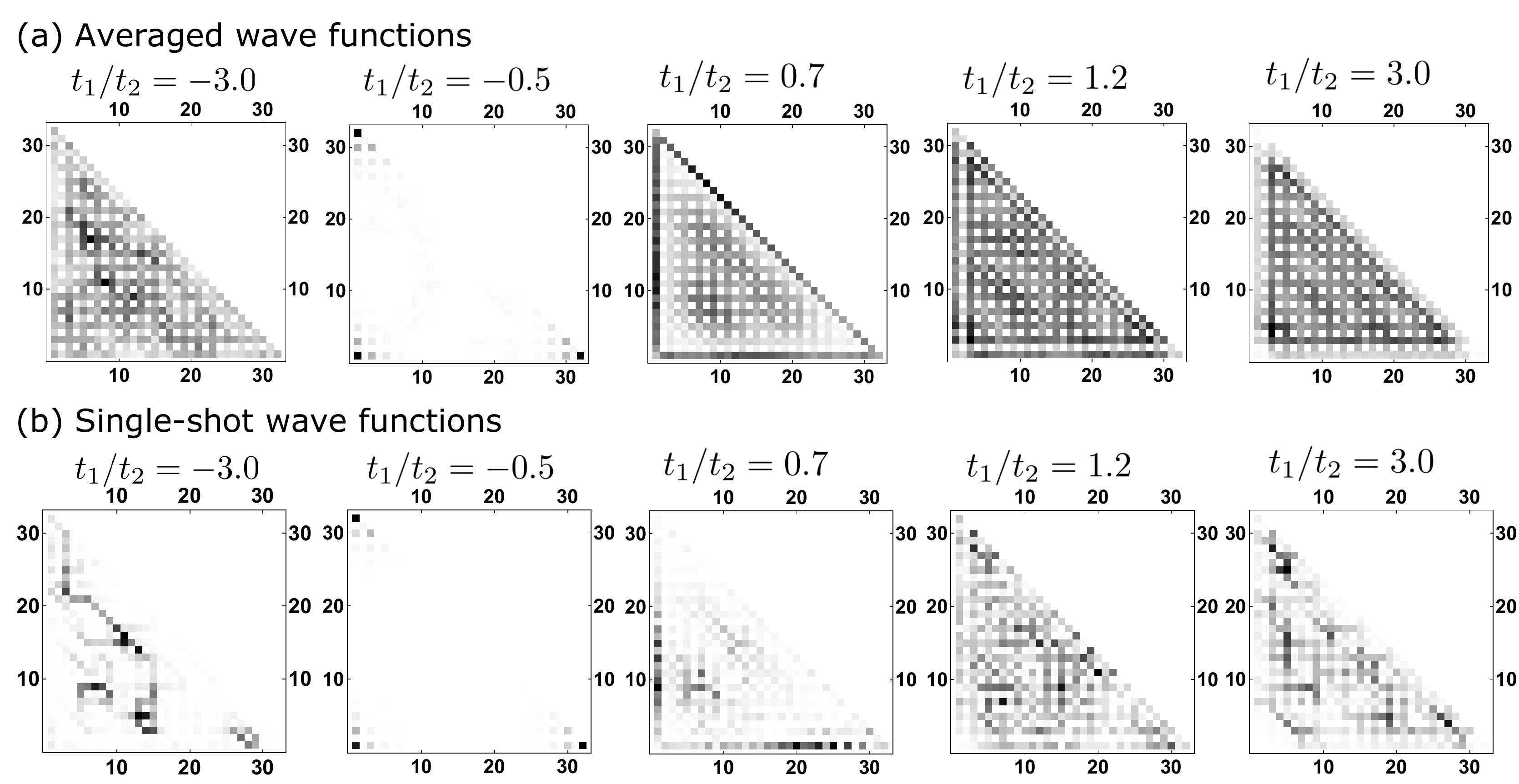}
\caption{
  Wave functions on a triangle geometry.
  (a) The averaged wave functions of the highest occupied state 
  with $W=1$.
  The average is taken over 30 samples.
  (b)
  The single-shot wave functions of the highest occupied state 
  with $W=1$.
  The average over three corner states is also taken (see the main text).
  }
  \label{fig:states_triangle}
 \end{center}
\end{figure*}

To clarify the difference between the averaged and the single-shot wave functions, we show 
the averaged wave functions at $W=1$ for several parameters in Fig.~\ref{fig:states}(a) (for a rhombus) and Fig.~\ref{fig:states_triangle} (a) (for a triangle),
and the single-shot wave functions at $W=1$ for the same parameters in Fig.~\ref{fig:states}(b) (for a rhombus) and Fig.~\ref{fig:states_triangle}(b)  (for a triangle).
Clearly, the restoration of the translational symmetry due to the averaging occurs at, for example, $t_1/t_2 = -3.0$, where the flat-band states are dominant. 
For other phases, the corner states exist in both the averaged and the single-shot wave functions, which serve as an indicator of the HOTI phase for the neural networks, as we will see later. 

\begin{figure*}[t]
 \begin{center}
  \begin{tabular}{c}

   \begin{minipage}{0.5\hsize}
   \begin{flushleft}
   {\Large (a) }
   \end{flushleft}
    \begin{center}
     \includegraphics[width=0.9\linewidth]{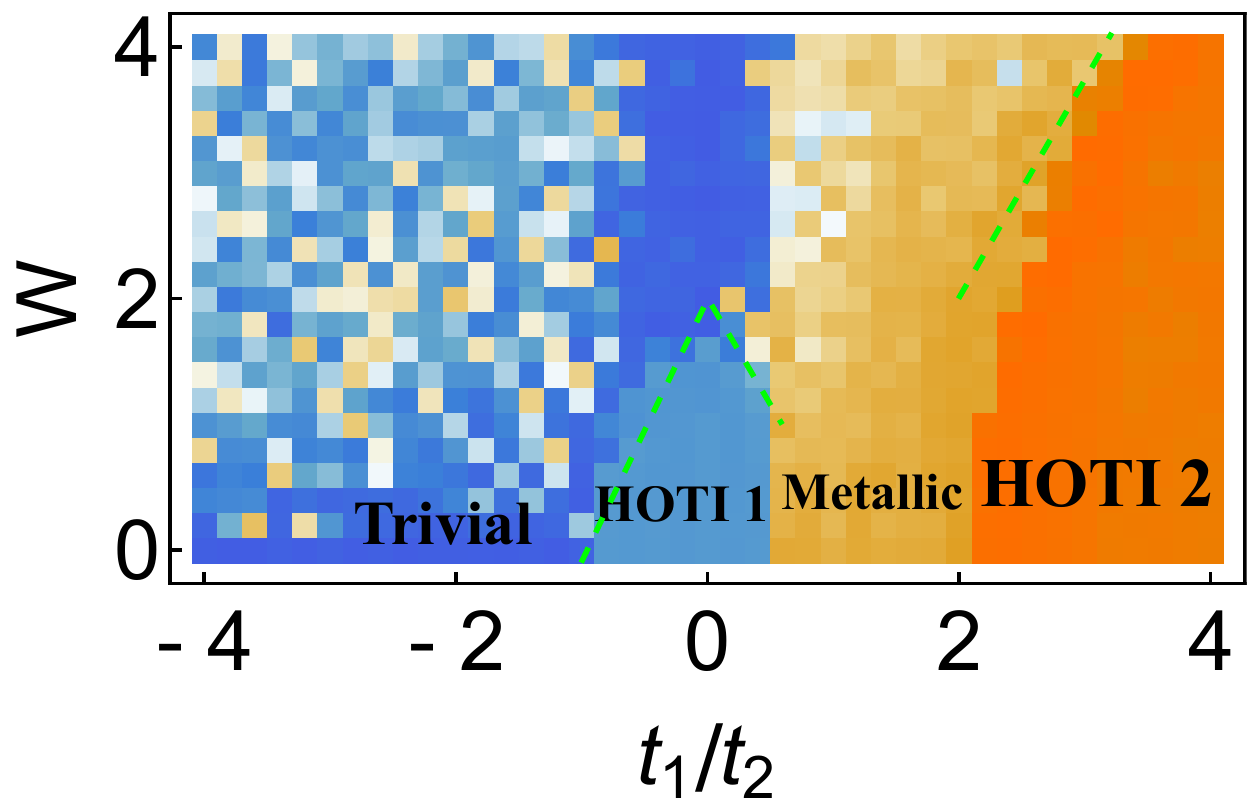}
    \end{center}
   \end{minipage} 

   \begin{minipage}{0.5\hsize}
   \begin{flushleft}
   {\Large (b) }
   \end{flushleft}
    \begin{center}
     \includegraphics[width=0.9\linewidth]{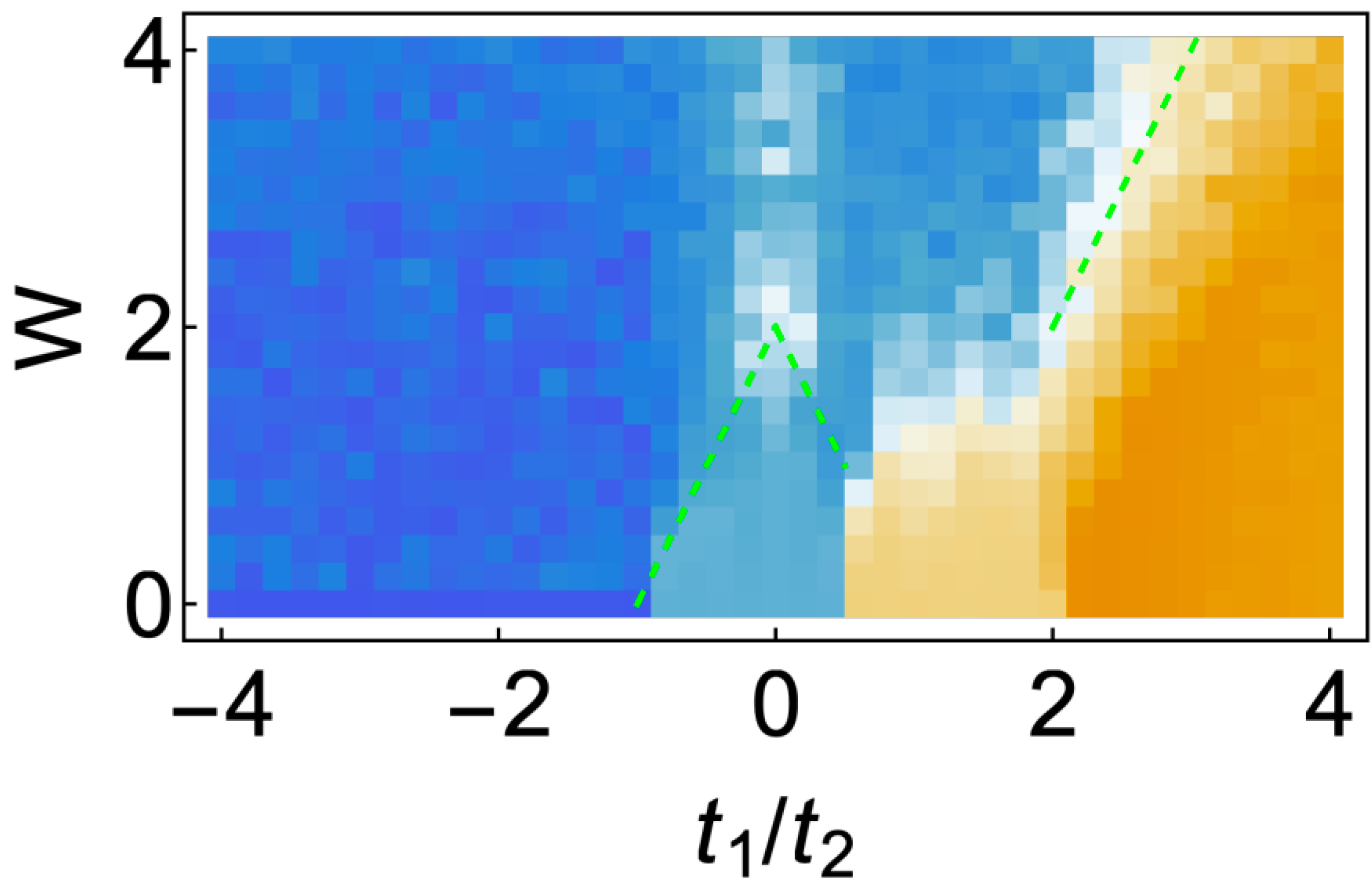}
    \end{center}
   \end{minipage} 
   
    \end{tabular}
  \caption{
  The phase diagram of the breathing kagome lattice with disorders for the rhombus geometry.
  (a) The phase diagrams is plotted for the averaged wave functions.
  The blue, right-blue, yellow and orange areas
  denote trivial, HOTI 1, HOTI 2 and metallic phases, respectively. 
  Green dashed lines denote the energy gap between the corner states and the bulk/edge states. 
  (b) The phase diagram is plotted by the averaged probabilities for 30 single-shot wave functions.
  }
  \label{fig:phase_diagram}
 \end{center}
\end{figure*}
\begin{figure*}[t]
 \begin{center}
  \begin{tabular}{c}

    \begin{minipage}{0.5\hsize}
    \begin{flushleft}
      {\Large   (a) }
     \end{flushleft}
    \begin{center}
     \includegraphics[clip, width=0.98\linewidth]{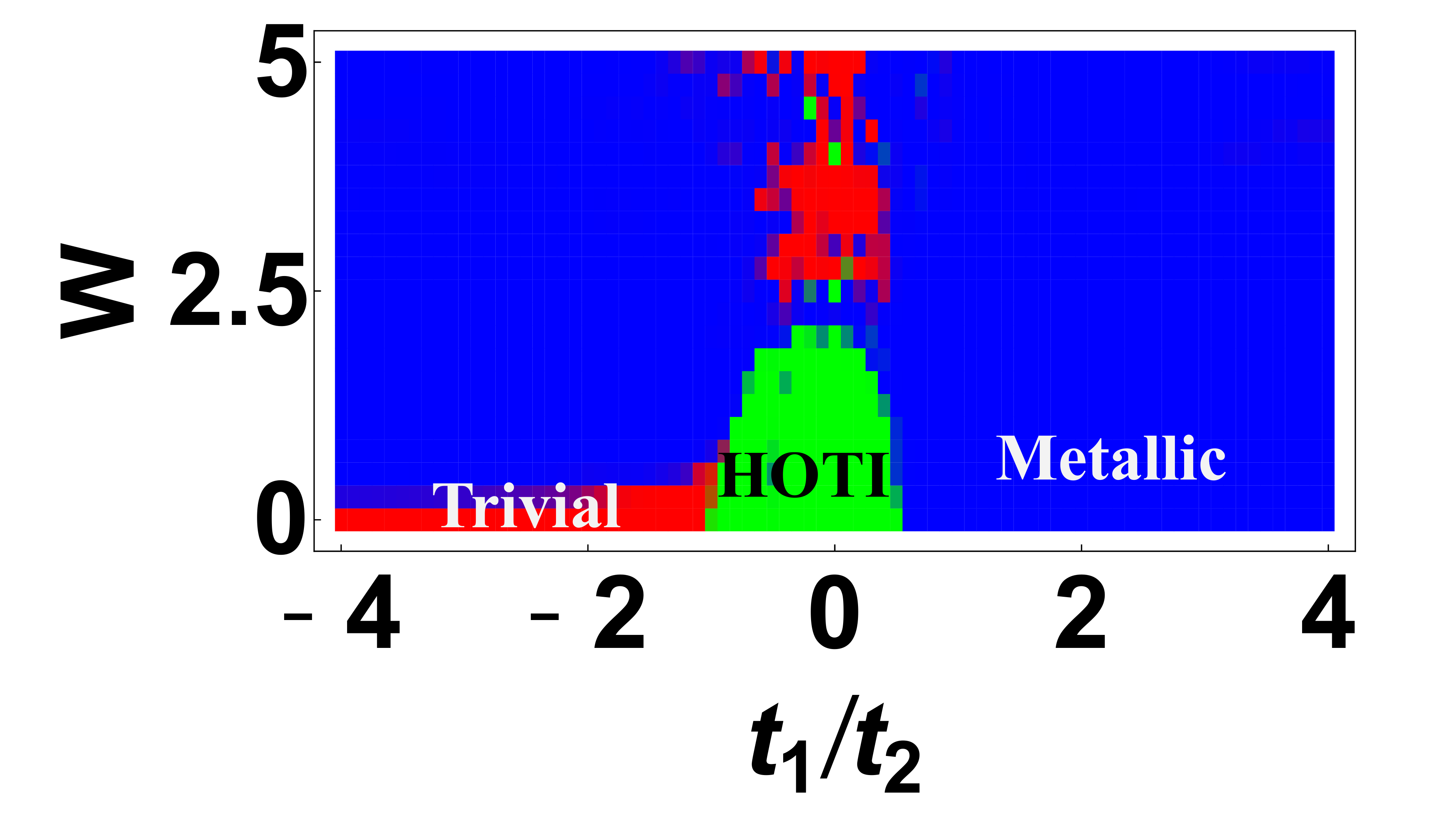}
    \end{center}
   \end{minipage}
   
    \begin{minipage}{0.5\hsize}
    \begin{flushleft}
      {\Large   (b) }
     \end{flushleft}
    \begin{center}
     \includegraphics[clip, width=0.98\linewidth]{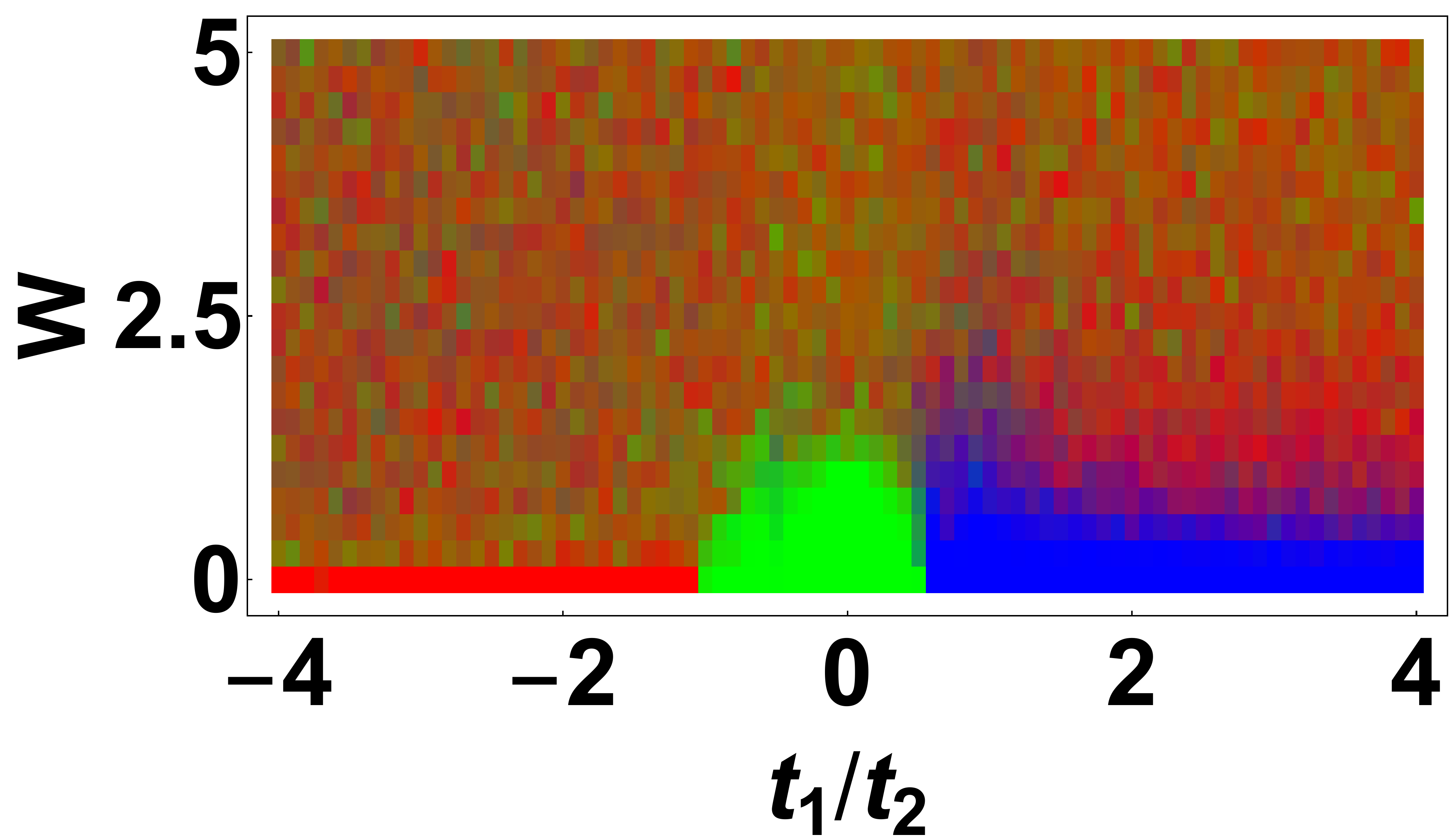}
    \end{center}
   \end{minipage}

    \end{tabular}
  \caption{
  The phase diagram of the breathing kagome lattice with disorders for the triangle geometry.
  (a) The phase diagrams is plotted for the averaged wave functions.
  The red, blue and green areas
  denote trivial, HOTI, and metallic phases, respectively.
  (b) The phase diagram is plotted by the averaged probabilities for 30 single-shot wave functions.
  }
  \label{fig:phase_diagram_triangle}
 \end{center}
\end{figure*}

\section{Results \label{sec:result} }
\subsection{Rhombus geometry}
We first consider the rhombus geometry with the system size $M=10$. 
The phase diagram for the averaged wave functions is shown in Fig. \ref{fig:phase_diagram}(a), and 
that for the averaged wave functions is shown in Fig. \ref{fig:phase_diagram}(b).
The colors represent the phases: Blue, right-blue, yellow, and orange regions, respectively, correspond to
trivial, HOTI 1, metallic, and HOTI 2 phases. 

First, let us focus on the HOTI phases. 
We see in both of two phase diagrams that the HOTI phases survive up to the critical strength of the disorders. 
The HOTI 1 phase turns into the trivial phase for $W > W^{(1)}_c$, 
while HOTI 2 phase into the gapless phase for $W > W^{(2)}_c$,
indicating that the level crossing between corner states and the states nearby zero-energy 
occurs in these regimes. 
If it is the case, one can estimate $W^{(1,2)}_c$ with respect to the gap 
between corner states and nearby bulk/edge states.
To be specific, the formulas for $W_c^{(1)}$ are $2 t_1/t_2 + 2 $ for $-1  \leq t_1/t_2  \leq 0 $, 
and $ -2 t_1/t_2 + 2 $ for $ 0 \leq t_1/t_2  \leq 1/2 $;
similarly, those for $W_c^{(2)}$ are $ 2 t_1/t_2  -2$ for $t_1/t_2 > 2 $.
These estimated values are denoted by green lines, which turn out to be good approximations.
Notice that this behavior is reminiscent of the Su-Schrieffer-Heeger model in the presence of on-site disorders, 
where the zero-dimensional edge states become unstable when the disorder strength is comparable with bulk band gap~\cite{Perez-Gonzalez2018,Munoz2018}.

Looking at the other phases, the trivial phase is the most fragile against disorders.
Indeed, we see in Fig. \ref{fig:phase_diagram}(a) that there is a mosaic-like region,
which means that the phase is ``unknown'' for the trained neural network.
To be more specific, in this region, the confidence is less than 98\% for all four phases.
We expect that this phase is likely to be an Anderson-localized (AL) phase, because of the following reason:
in this regime, the highest occupied state belongs to a flat band, which has a massive degeneracy. 
Then, the disorders lift the degeneracy and pick one of the localized state.
Qualitative argument on this localization transition by using the inverse participation ratio (IPR) is presented in Appendix.
We remark that the previous study predict the existence of the critical phase between the trivial and the Anderson-localized phase
~\cite{PhysRevB.82.104209},
but the transition occurs at very small disorders ($W < 10^{-2}$), 
which is much smaller than our minimum strength of the disorder ($W = 0.2$).
Also, the re-entrance to the metallic phase predicted in Refs. \onlinecite{PhysRevLett.96.126401,doi:10.1143/JPSJ.76.024709} occurs at much larger $W$ than the maximum value we study.
 
\subsection{Triangle geometry}
Next, we show the results for the triangle geometry. 
The phase diagrams for $M=16$ are shown in Fig. \ref{fig:phase_diagram_triangle}. 
Here, the output in the form of three-dimensional vectors is mapped onto the RGB component of the color map; 
red, blue, and green regions respectively, correspond to the trivial, HOTI, and metallic phases. 

We see that the HOTI phase 
vanishes above a critical strength of the disorder potential, which is again roughly estimated by the energy gap. 
This behavior is similar to the rhombus geometry, so
one may speculate that the HOTI phase is stable against the disorders as far as the corner states survive.

We also see that there is a metallic phase above the trivial phase for the result of the averaged wave function [Fig. \ref{fig:phase_diagram_triangle}(a)].
This is the artifact of the averaging of the wave functions, as we described in the previous section. Indeed, 
we clearly see the mosaic-like pattern in the result for the single-shot wave function [Fig. \ref{fig:phase_diagram_triangle}(b)],
clearly indicating the existence of the AL phase: this is consistent with the observation from the wave function (Fig. \ref{fig:states_triangle}).
Further, we also observe that the Anderson localization occurs above the metallic phase, 
and that its critical value is smaller compared with the case of the rhombus geometry.

\section{Summary and discussions \label{sec:summary}}
To summarize, we have investigated the phase diagram of the disordered breathing kagome model by using a machine learning method.
We use the wave function of the highest occupied state as input data.
By doing so, the neural network can distinguish the HOTI phase from other phases by the existence of the corner states.
Our results reveal that the HOTI phase is robust against the disorders as far as the disorder strength does not exceed the energy gap. 
Besides the HOTI phase, the AL state and the trivial/metallic states can be successfully distinguished by combining 
the results of averaged wave functions and of the single-shot wave functions.

It is important to justify our results of the machine learning method by using conventional approaches to study the disordered models. 
As an example of such conventional approaches, we investigate the IPR for the present model with the rhombus geometry, which provides a consistent result
with that of the machine learning (see. Appendix). 
Testing various other approaches, such as the transfer matrix method~\cite{PhysRevLett.110.236803,doi:10.7566/JPSJ.85.123706, doi:10.7566/JPSJ.86.044708,PhysRevB.97.205110,MacKinnon1983}, 
will be an interesting future problem. 

Finally, let us remark on the related works. 
Recently, the robustness of the corner states against disorders was investigated
in the electric circuits~\cite{Imhof2018, PhysRevB.98.201402}. 
There, it was found that the corner-mode-induced resonance peak of the impedance survives even though the capacitances and inductances are disordered slightly.
Although disorders of this kind in the electric circuits induce both on-site and bond disorders in the language of the tight-binding model~\cite{Lee2018}, 
these results on the robustness of the corner states are consistent with our present results.

\acknowledgments
The authors would like to thank Nobuyuki Yoshioka for drawing our attention to 
application of machine learning techniques to topological phases. 
They also thank Yutaka Akagi and Hosho Katsura for the useful comments.
This work is partly supported by JSPS KAKENHI Grants No. JP17H06138 and JP16K13845.

\appendix*

\section{Inverse participation ratio \label{app:b}}

\begin{figure*}[htb]

  \begin{center}
   \includegraphics[width=1.0\linewidth]{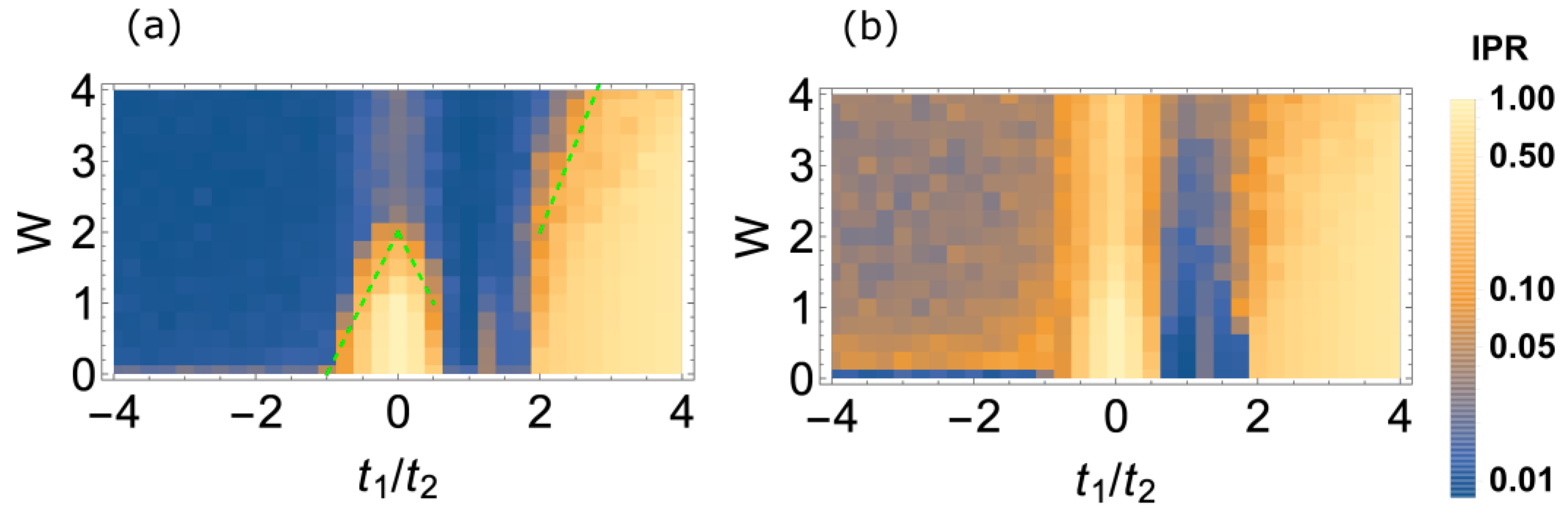}
  \end{center}

  \caption{
  (a) The IPR for the averaged wave functions. (b) The averaged IPR for the single-shot wave functions.
  The number of the samples are 30 for each parameters.
  The system size $M=10$. 
  Green dashed lines are same to the lines in Fig. \ref{fig:phase_diagram}.
  These figures are plotted on a logarithmic scale.
  }
  \label{fig:L4 norm}

\end{figure*}
The IPR is a useful quantity to distinguish the localized state from the extended state~\cite{PhysRevB.31.6146, PhysRevA.33.1141, doi:10.1143/JPSJ.43.415, doi:10.1143/JPSJ.76.024709}.
The IPR is defined as 
\begin{equation}
 p = \sum_{i}|\phi(r_i)|^4.
\end{equation}
Here, the eigenstate is normalized ($\sum_{i}|\phi(r_i)|^2 = 1$).
For the fully extended states, $\phi(r_i) = 1/\sqrt{N}$ hence $p \simeq 1/N$, 
where $N$ is the number of sites. 
So the IPR for the extended states vanishes in the
large-system-size limit.
On the other hand, IPR for the AL states takes larger value and not vanished in the
large-system-size limit. 

For the breathing kagome model, the corner states and the AL states
are the localized states and trivial and metallic states are the extended states,
so the IPR for the single-shot wave function will distinguish these two classes of phase.  
Further, to distinguish the corner state and the AL states, 
we need to calculate the IPR for the averaged wave function and compare it with the result for the single-shot wave function. 
Namely, when the AL states is averaged over a number of samples, 
the distribution of the wave function becomes uniforms and the IPR thus obtained is vanishing, as is in the case of extended states.
On the other hand, for the corner states, the distribution is still weighted at the corners even for the averaged wave functions,
(see Figs. \ref{fig:states} and \ref{fig:states_triangle}), thus the IPR is not vanishing. 
The expected behaviors for each phase is summarized in Table \ref{Table1}.

The results of the IPR for a rhombus geometry is shown in 
Fig. \ref{fig:L4 norm}(a) for the averaged wave functions, and Fig. \ref{fig:L4 norm}(b) for the single-shot wave function. 
Here the number of samples is 30.
Comparing the results with the phase diagram obtained by the machine learning (Fig. \ref{fig:phase_diagram}),
we see that the IPR indeed shows the expected behaviors as in Table \ref{Table1},
indicating the reliability of the machine learning method. 

Besides the correspondence between the IPR and the phase diagrams from the machine learning, 
we also find that the large IPR is obtained at $t_1/t_2  = 1$, where the Fermi level is at the Dirac point in the clean limit, for weak $W$. 
This may be attributed to strongly-localized zero-energy modes inherent in the disordered Dirac fermion systems~\cite{PhysRevB.48.4204}.

\begin{table}[b]
  \caption{The expected behaviors of the IPR for HOTI, AL, and trivial/metallic phases. }
  \label{Table1}
\begin{center}
  \begin{tabular}{c|c|c|c|}
  & HOTI & AL & Trivial/metallic \\ \hline
  Single-shot & $\sim 1$ & $\sim 1$ & $\sim 0$\\
  Averaged& $\sim 1$ & $\sim 0$ & $\sim 0$\\
     \end{tabular} 
\end{center}
\end{table}

%

\end{document}